\documentclass[aps,prl,preprint,amsmath,amssymb,nofootinbib,unsortedaddress]{revtex4-1}

\usepackage{braket} 
\usepackage{amsmath,amssymb,bm,graphicx,epsfig,psfrag,ulem,color,natbib}
\usepackage{units}
\usepackage{float} 
\usepackage{amsmath}   
\usepackage{upgreek}  
\usepackage{bm}  
\usepackage[latin1]{inputenc}
\usepackage{epstopdf}
\newcommand{\boldr}{{\bf{r}}}

\begin{document}

\title{Twisted unwinding of multi-charged quantum vortex \\
and generation of turbulence in an atomic Bose--Einstein condensate}

\author{G.~D.~Telles, P.~E.~S.~Tavares, A.~R.~Fritsch, 
A.~Cidrim, V.~S.~Bagnato}
\affiliation{Instituto de F\'{\i}sica de S\~{a}o Carlos,
Universidade de S\~{a}o Paulo, C.P. 369,
13560-970 S\~{a}o Carlos, SP, Brazil}

\author{A.~C. White}
\affiliation{Quantum Systems Unit, Okinawa Institute of Science and
Technology, Okinawa 904-0495, Japan}

\author{A.~J.~Allen, C.~F.~Barenghi}
\affiliation{Joint Quantum Centre (JQC) Durham-Newcastle,
School of Mathematics and Statistics,
Newcastle University, Newcastle upon Tyne NE1 7RU, England, UK}

\begin{abstract}
We report the observation of the twisted decay 
of quadruply charged vortices in an atomic 
Bose-Einstein condensate. Supporting numerical simulations show that the singly-charged vortices, which result from the decay of a multi-charged vortex, twist around intertwined in the shape of helical Kelvin waves. Finally, we propose to apply this effect to generate an almost isotropic state of turbulence which we characterize in terms of the velocity statistics.

\end{abstract}


\maketitle

Superfluids are noteworthy because they flow
without dissipating energy. Even more remarkably, superfluid flow patterns
are characterized by the quantization of vorticity, arising
from the existence and the uniqueness of a macroscopic 
wave function $\psi$. The flow velocity is proportional
to the the phase gradient of $\psi$, and the circulation around a vortex 
line must be an integer multiple $n=1,2,\cdots$ of the quantum of circulation $\kappa=h/m$, where $h$ is Planck's constant and $m$ is the atomic mass. The singular nature of this quantized vorticity, concentrated along lines, has yet another important consequence: within a turbulent tangle of quantum
vortices, the velocity components obey power-law statistics
\cite{Paoletti2008,White2010}, unlike Gaussian statistics typical of 
ordinary turbulence. Recent experiments and numerical simulations \cite{Barenghi2014b} have shown that, under certain conditions, the turbulent superflows share a remarkable property with classical turbulence: the same Kolmogorov energy spectrum \cite{Frisch},
describing the distribution of kinetic energy over the large 
length scales. This property suggests that quantum turbulence may be
the `skeleton' of classical turbulence \cite{Hanninen2014}.

Atomic Bose-Einstein condensates (BECs) are emerging as ideal systems 
to explore the quantization of vorticity and many other fundamental problems 
concerning the nature of turbulence \cite{Tsatsos2016}. Vortices are more 
easily nucleated, manipulated \cite{Aioi2011,Davis2009}
and observed \cite{Madison2000,Raman2001,Freilich2010}
in BECs than in superfluid helium, due to the typical vortex core sizes 
which are orders of magnitude larger in gaseous BECs 
($\approx 10^{-7}~{\rm m}$) than in liquid He ($\approx 10^{-10}~{\rm m}$).

Multiply quantized vortices are interesting especially since they are energetically unstable and decay into singly quantized vortices~\cite{Shin2004,Kumakura2006,Okano2006,Isoshima2007}. The angular momentum and the energy of an isolated vortex in a homogeneous
superfluid grow respectively with $n$ and $n^2$~\cite{Donnelly1991}. 
For the same angular momentum, multi-charged ($n>1$) vortices carry more energy, and, in the presence of dissipative mechanisms (e.g. thermal excitations), tend to decay into singly-charged vortices, 
minimizing the system's energy. Besides the energy instability, there is also a dynamical instability~\cite{Pu1999,Mottonen2003,Huhtamaki2006a}, which is able to destabilize the
vortices even in the absence of dissipation, at zero temperature, whose time-scales have been investigated 
\cite{Mottonen2003,Shin2004,Huhtamaki2006b,Mateo2006,Okano2006,Isoshima2007,Kuwamoto2010,Kuopanportti2010}.
The technique of topological phase imprinting \cite{Nakahara2000}
has allowed the controlled generation of multi-charged vortices
\cite{Leanhardt2002,Shin2004,Okano2006,Kuwamoto2010} in atomic condensates. 
The splitting of doubly quantized vortices into two singly quantized vortices 
has been observed in a Na BEC and the decay process was 
studied~\cite{Shin2004,Huhtamaki2006b,Mateo2006}. 
Quadruply quantized vortices are theoretically 
predicted~\cite{Kawaguchi2004} to decay presenting various symmetries 
of splitting, making them an interesting research topic. 
Recent work has determined that the 
stability of such vortices is affected by the condensate's 
density \cite{Shin2004} and size \cite{Mateo2006}, and by the
nature of the perturbations \cite{Kawaguchi2004}.

In this letter, we discuss the decay of an $n=4$ charged vortex line  
nucleated in prolate $^{87}\rm{Rb}$ BECs. By combining experimental 
observations with numerical simulations, we present results 
showing the twisted decay process of a multi-charged vortex, 
induces helical Kelvin waves on the resulting singly-charged vortex lines. 
We also show that the observed intertwined decay of multi-charged vortices 
may be exploited to create an almost isotropic state of quantum turbulence 
in atomic condensates. Kelvin waves are of particular interest because 
they are thought to play a key role in quantum 
turbulence \cite{Barenghi2014a}, being observed in classical fluids and 
superfluid helium, including atomic condensates \cite{Bretin03,Smith04}.

\begin{figure}[!ht]
\centering{
\includegraphics[scale = 0.45]{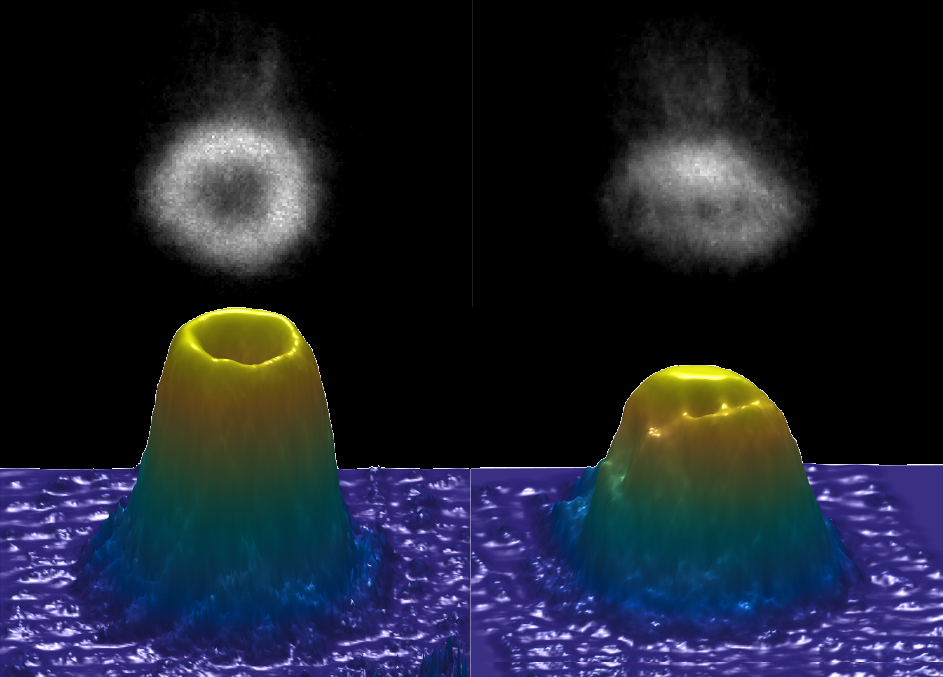}}
\caption{{\bf Top:}
Image showing the $n=4$ multi--charged vortex under free fall(left) and
the four singly-charged vortices resulting from the $n=4$ vortex decay
(right). Both top images present the integrated number density 
acquired along the BEC $z$-axis, after $t=14~\unit{ms}$ of TOF.
{\bf Bottom:} corresponding 3D representation of the
top right and left images of the optical densities to emphasize the vortex core depletions. 
Typical optical density linear profiles used for data analysis are
displayed in Fig.~2 of the Supp. Mat. \cite{Supp2}.
}
\label{fig1}
\end{figure}

\begin{figure}[!ht]
\begin{minipage}[t]{0.6\textwidth}
  \centering
  \includegraphics[scale = 0.48]{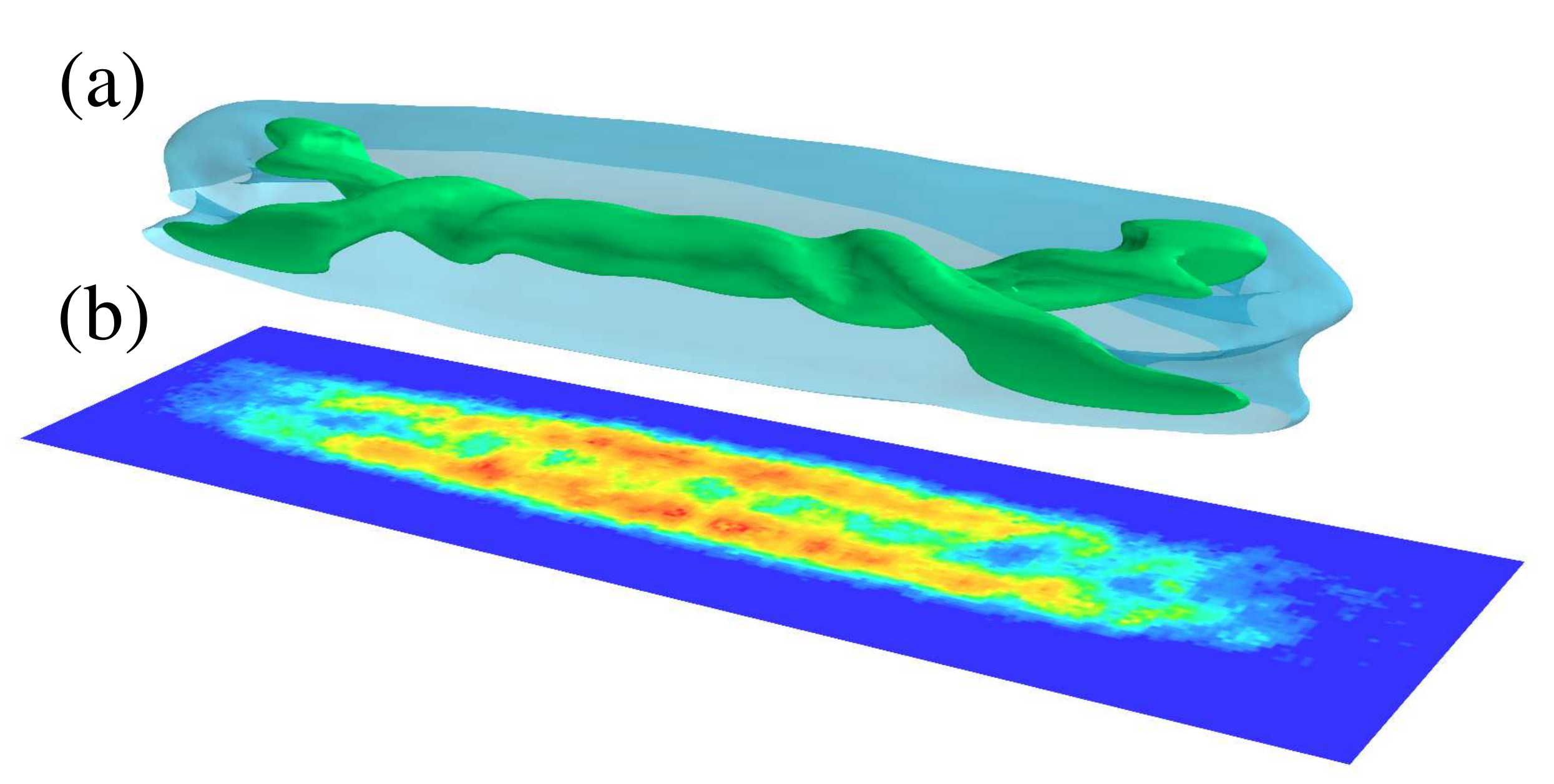}
  \includegraphics[scale = 0.55]{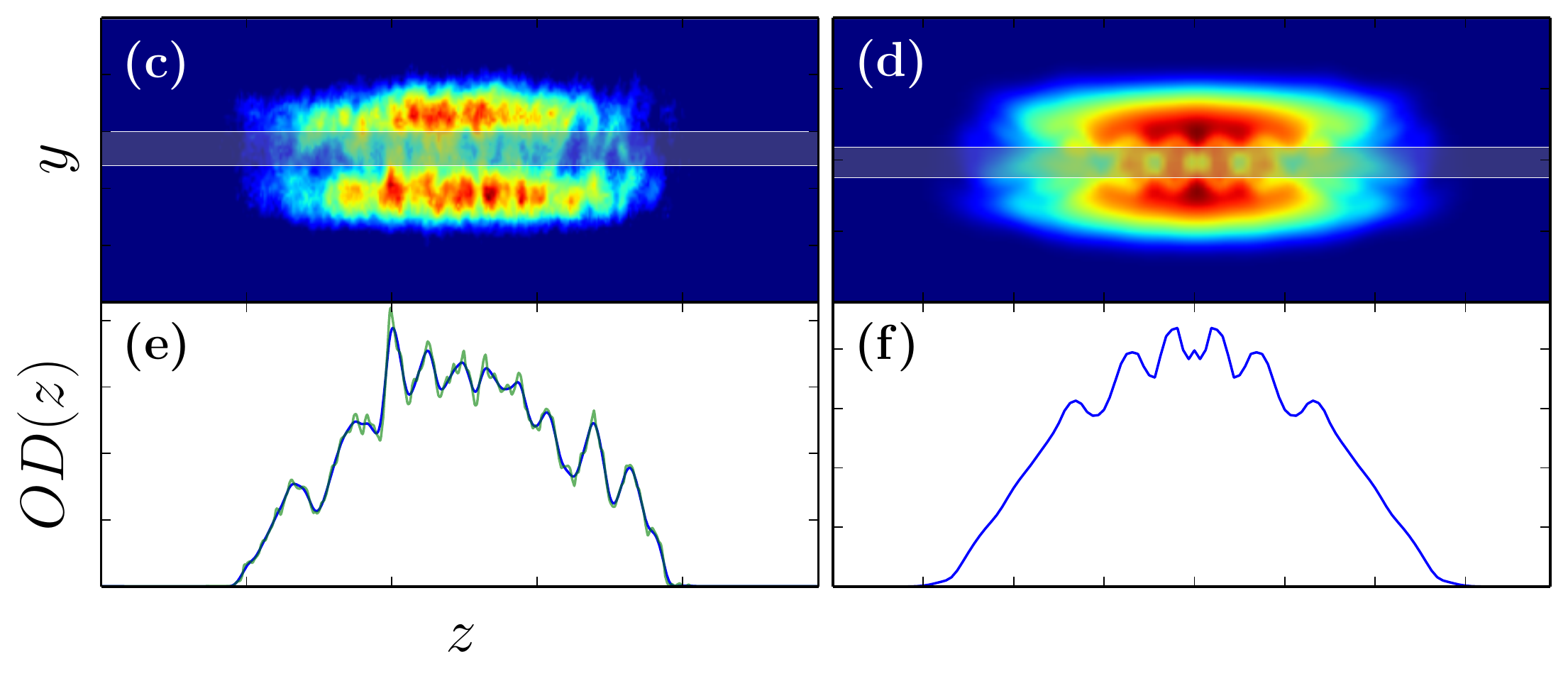}
\end{minipage}
\caption{{\bf Top:} (a) the 3D isosurface resulting from the GP simulation solved for the same experimental conditions. (b) side absorption image showing central density oscillations resulting from the twisted unwinding decay. {\bf Center:} (c) side experimental absorption image, also shown in (b), presented in false color alongside its corresponding (d) 2D column density resulting from the GP simulation. {\bf Bottom:} (e) experimental $OD$ $z$-profile, and the corresponding (f) numerical central $OD(z)$ along the condensate $z$-axis (averaged over the narrow vertical gap in between the horizontal strips shown). In (e) The smoothed (blue) curve plotted on top the raw experimental $OD$ (green) curve after application of a high-frequency filter.}
\label{fig2}
\end{figure}

The experimental sequence to produce the BEC runs as follows. First, $^{87}\rm{Rb}$ BECs are produced in the $\left|F=2,m_{F}=+2\right>$ hyperfine state, with a small thermal fraction in a cigar-shaped  QUIC magnetic trap. The atoms undergo forced evaporation for $\unit[23]{s}$, from $\unit[20]{MHz}$ to about $\unit[1.71]{MHz}$, following a non analytic curve, experimentally determined by optimizing the phase space density and the runaway condition in each step. We typically produce samples with no more than $35\%$ of thermal fraction, well into the Thomas-Fermi regime. The measured trapping frequencies are: $\omega_z=2\pi \times 21.1(1)\unit{Hz}$ in the symmetry axis, and $\omega_r=2\pi \times 188.2(3)\unit{Hz}$ in the radial direction, resulting in the (geometric) mean trap frequency, $\bar{\omega}=2\pi \times 90.7(4)\unit{Hz}$, and the harmonic oscillator length, $a_\text{ho}=\sqrt{\hbar/m\bar{\omega}}\approx \unit[1.13]{\mu m}$. 
The typical properties of our BECs are: reduced
temperature $T/T_0 \approx 0.65 - 0.70$; $N_{0}\approx 1-2~\times 10^{5}$ condensate atoms; $N \approx 1-3 ~\times 10^5$ 
total number of atoms; healing length $\xi \approx \unit[0.17]{\mu m}$; 
axial and radial Thomas-Fermi radii 
$R_{TF}(z,0) = \unit[13.9(7)]{\mu m}$ and
$R_{TF}(r,0) = \unit[5.8(6)]{\mu m}$, respectively; 
chemical potential $\mu \approx \unit[100]{nK}$; 
anisotropy parameter $\lambda={\omega_z}/{\omega_r} \approx 0.112$.
The multi-charged vortices are nucleated by adiabatically
reducing and inverting the initial trap bias field from 
$B_{0}(t=0)\approx \unit[0.5]{G}$ 
down to $B_{0}(t=\tau)\approx \unit[-0.5]{G}$, 
in $\tau\approx \unit[5]{ms}$ typically. The resulting magnetic 
bias field along the weak trap direction, $z$, is reversed during 
the process \cite{Supp1}. 
We find that the $n=4$ charged vortices, topologically 
imprinted, are unstable and decay into four singly-charged vortices 
in a twisted unwinding manner. The process is very reproducible, 
over several runs and different time-of-flight values. 
The 2D optical depth (OD) images presented correspond to the 
integrated number density acquired along the symmetry axis (axial images), 
shown
in the top of Fig.~\ref{fig1}, or integrated along $y$ (side images),  
shown from the side of the condensate, see Figs.~\ref{fig2}$b$, 
and \ref{fig2}$c$. The experimental images are carefully compared to 
the corresponding 2D density maps (Fig.~\ref{fig2}$b$) and 3D 
isosurfaces (Fig.~\ref{fig2}$a$) resulting from the numerical simulations.

Given the relatively low condensate temperatures, we model the 
condensate's dynamics using the Gross-Pitaevskii equation 
(GPE) \cite{PitaevskiiStringari2003} 
\begin{equation}
       i \hbar \frac {\partial \psi}{\partial t} = 
       \left( 
       -\frac{\hbar ^2}{2m}\nabla^2 + V + g\vert \psi \vert^2  
       \right) \psi,
\label{eq:gpe}
\end{equation}

\noindent
where $\psi(\boldr,t)$ is the condensate's wavefunction, 
$\boldr$ the position, $t$ the time, 
$V(\boldr)=m(\omega_x^2 x^2 + \omega_y^2 y^2 + \omega_z^2 z^2)/2=
m(\omega_rr^2+\omega_zz^2)/2$
the trapping potential,
$g=4\pi \hbar^2 a_s/m$ the strength of the inter-atomic interactions and
$a_s$ the s-wave scattering length. 
The normalization is $\int_V \vert \psi \vert^2 dV=N$ where
$V$ is the condensate volume. 
The GPE is cast in dimensionless form using harmonic oscillator
units and solved numerically using the 4th order Runge Kutta method with the help of XMDS \cite{Dennis2013}. We simulate the 
vortex decay for a BEC cloud with $N\approx 1\times 10^{5}$ atoms 
and the same radial and axial harmonic trap frequencies of
the experiment.

The top left picture in Fig.~\ref{fig1} shows a typical BEC axial optical depth image of a $n=4$ charged vortex before it decays; on the bottom left, the corresponding number density surface.
Note the large central circular depletion, representing its core. The top right and bottom images show a BEC containing four singly-charged vortices nucleated by the twisted decay of the initial quadruply charged vortex.

Fig.~\ref{fig2} (a and b) show the computed 3D isodensity surface (\ref{fig2}a) on top of its corresponding experimental absorption image (\ref{fig2}b), respectively. Note how the vortex lines appear intertwined, in the form of helical Kelvin waves along the $z$-axis, as predicted~\cite{Mottonen2003,Huhtamaki2006b,Mateo2006}. We interpret the central region of Fig.~\ref{fig2}b, containing a nearly regular density modulation along the symmetry axis, as a signature of the presence of Kelvin waves.
Fig.~\ref{fig2} (c and d) are shown, allowing for a direct comparison of the similar features appearing on the experimental (Fig.~\ref{fig2}c) and the numerical (Fig.~\ref{fig2}d) column density images. 
Moreover, we present experimental (Fig.~\ref{fig2}e) and numerical 
(Fig.~\ref{fig2}f) optical density profiles, OD(z), directly 
extracted from the central (shaded) regions of Figs.~\ref{fig2}c 
and \ref{fig2}d, along the $z$-axis.
The regular oscillations observed near the center on the OD(z) profile
demonstrate that the density modulations are due to the presence of 
Kelvin waves, clearly visible in the numerical 3D isosurface 
(Fig.~\ref{fig2}a). Likewise, the observation of Kelvin waves reported 
by Ref.~\cite{Bretin03} was supported by similar evidence. 
Without Kelvin waves, the central regions of the 2D column density 
images would be smooth, and the 1D OD(z) profiles would not present 
the oscillations  shown in Figs.~\ref{fig2}(e and f), as
demostrated in Fig.~3 of the Supp. Mat.~\cite{Supp3}.

The shape of the vortex lines resulting from the decay of multi-charged vortices depends on where and when the decay starts. They may appear as straight vortex lines or intertwined, as here reported, depending on the perturbation's symmetry and the local density homogeneity.
If the perturbation is mostly constant, along $z$, and the density does not
vary much in the $z$-direction, every point on the vortex unwinds
at the same rate, and singly-charged vortex straight lines are expected to emerge.
However, if the density changes significantly along $z$, the unwinding takes place at different times and at different $z$ positions, inducing the intertwining, as discussed by \cite{Huhtamaki2006b}.

\begin{figure}[!ht]
\centering{\includegraphics[scale = 0.55]{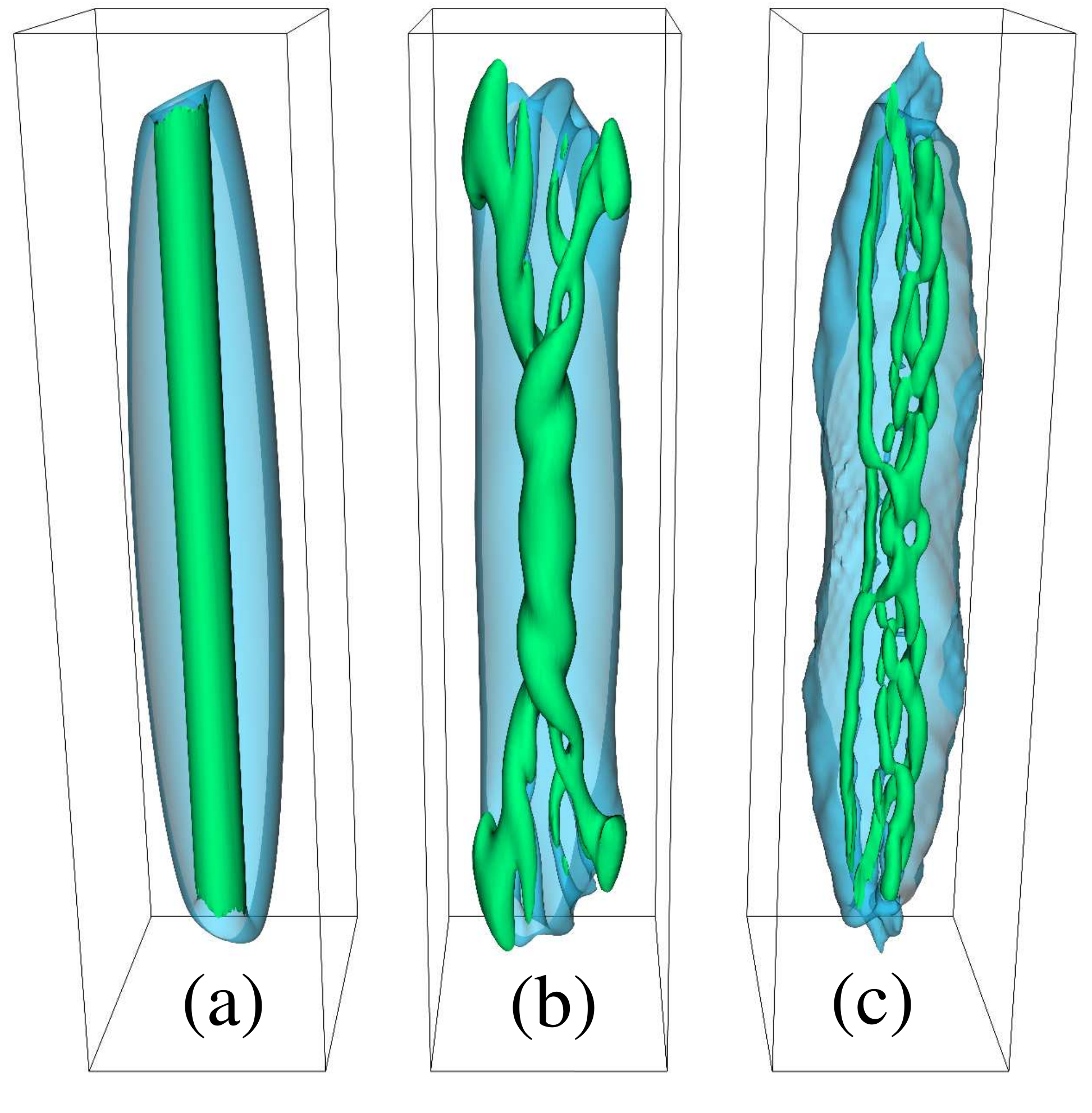}}
\caption{Numerically computed time sequence of 3D isodensity plots showing 
the initial $n=4$ multi-charged vortex (a) evolves to the twisted unwinding (b), and finally decays to four singly-charged vortices (c).
}
\label{fig3}
\end{figure}

The complete time sequence of the twisted vortex decay is revealed 
by the numerical simulations, as shown in the movie~\cite{SuppMovie}. 
Fig.~(\ref{fig3}) presents the three key features 
(the multi-charged vortex, the twisted unwinding, the four singly-charged
vortices) observed during the intertwined decay at different stages. 
It is worth comparing this effect, in which helical Kelvin waves arise
from the interaction of parallel vortices in a confined geometry, with
the Crow instability \cite{Simula2011}, which generates Kelvin waves
on anti-parallel vortices. Individual helical Kelvin waves have been recently observed in superfluid helium following a vortex reconnection \cite{Fonda2014}, whereas in our case, the waves arise from the decay of multiply charged vortices, as predicted by \cite{Mottonen2003,Mateo2006}.  
  
Numerical experiments suggest that the decay start 
of the multi-charged vortex can be sped up. Imposing 
random fluctuations ($\leqslant 10\%$ of $\vert \psi \vert$) to the 
initial $n=4$ vortex wave function does not significantly change the decay time scale, probably because the symmetry of the initial condition is not completely broken.  A small displacement of the vortex core axis 
($\approx a_{\text{ho}}/5$) is more efficient, triggering the onset of the twisted unwinding in about \unit[12]{ms}; a larger displacement
 ($\approx  a_{\text{ho}}/2$) reduces this time to \unit[10]{ms}. A few other methods were investigated, and the most efficient method found was to gently squeeze the harmonic potential in the $xy$ direction by an amount $\omega_x/\omega_y=0.9$, when preparing the initial state in imaginary time, then resetting $\omega_x/\omega_y=1$ while propagating the GPE in real time; this triggers the onset of decay in \unit[5]{ms}.
Experimentally, we found it difficult to control perturbations well enough to reproducibly determine the time scale of decay. It was observed that, once in inverted bias field configuration, the small instabilities existing on the current controller were sufficient to trigger in just \unit[5]{ms}, which also corresponds to the hold time used for taking the data presented.
\begin{figure}[!ht]
\centering{\includegraphics[scale = 0.55]{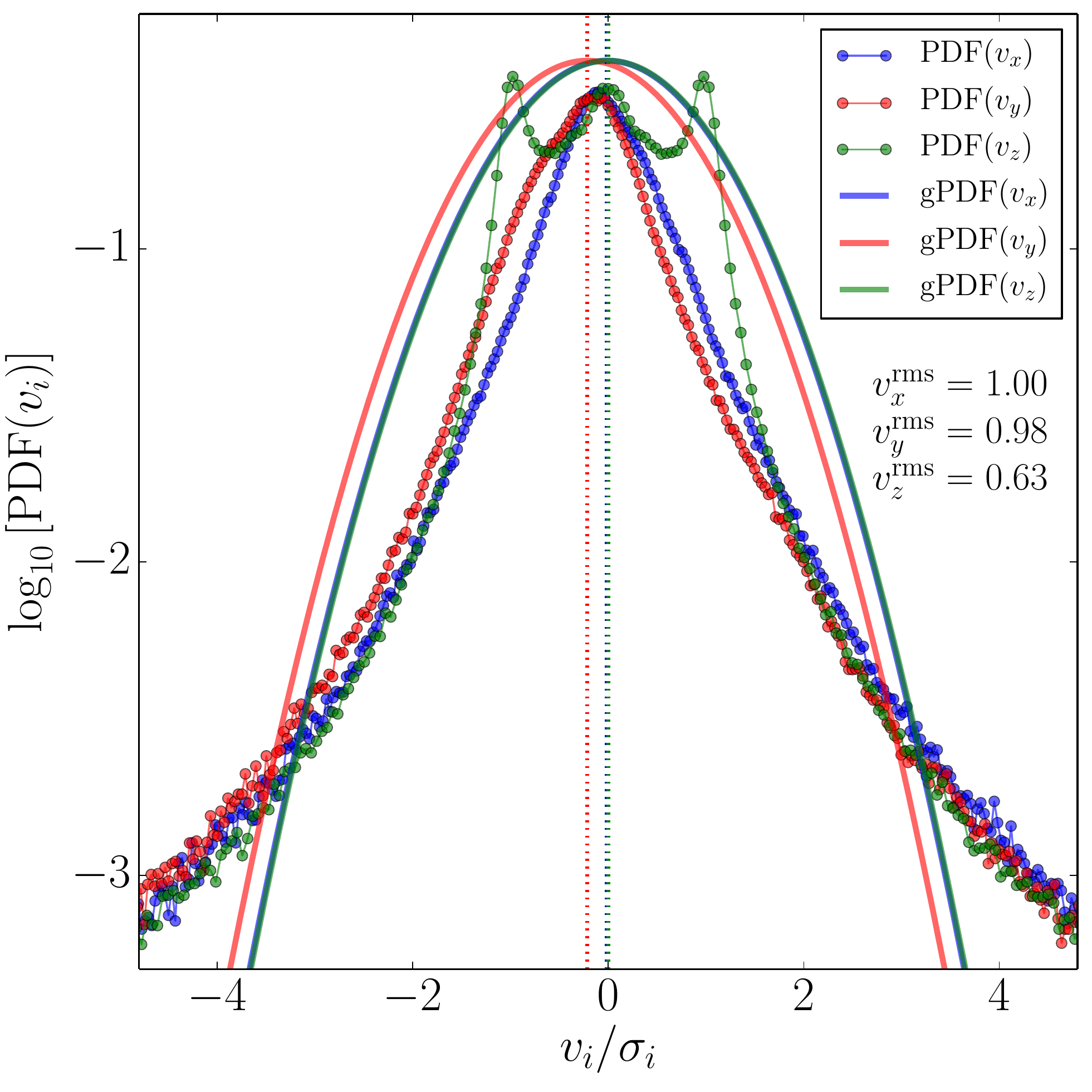}}\\
\caption{Turbulent decay of two antiparallel $n=2$ vortices.
PDFs of turbulent velocity components (blue, red, and green symbols
respectively for $x$, $y$, and $z$ components) at $t=\unit[53]{ms}$. Note
the power law nature of the PDFs at large velocity, unlike the
Gaussian fits (gPDFs), which are plotted as solid lines.
}
\label{fig4}
\end{figure}

As a final remark, it is worth mentioning that
the observed twisted decay of multi-charged vortices may 
be exploited to induce quantum turbulence. The existing methods 
used to generate turbulence (rotations~\cite{Kobayashi2007}, 
trap oscillations~\cite{Henn2009}, 
moving obstacle~\cite{Raman2001,Neely2010,White2012,Kwon2014,Stagg2015,
White2014,Allen2014b,Cidrim2016}, and imprinting staggered 
vortices~\cite{White2010}) tend to significantly perturb 
or even fragment\cite{Parker2005} the condensate. 
This complicates detecting the vortices position and counting 
their number, which affects the vortex line density estimation. 
Here we numerically imprint
two antiparallel doubly quantized ($n=2$) vortices in the $xy$ plane.
The vortices unwind and twist, moving slightly forward due to
their self-induced velocity, then the helical waves travelling 
in opposite directions reconnect, 
generating a turbulent tangle with only moderate density oscillations
\cite{Supp4}.
We find that the condensate's velocity field, initially anisotropic
($v_y/v_x \approx 1$ and $v_z/v_x \approx 0$), becomes almost isotropic in the turbulent stage ($v_y/v_x \approx 0.98$ and $v_z \approx 0.63$), 
displaying velocity PDFs with the typical power-law scaling at high velocity (see Fig.~\ref{fig4}) observed in superfluid helium \cite{Paoletti2008} and in larger, initially isotropic condensates \cite{White2010} (whereas in ordinary turbulence such PDFs are Gaussian).

In conclusion, by carefully comparing experimental and numerical
results, we have demonstrated that the decay of multiply charged
vortices, topologically imprinted in trapped atomic BECs, generate intertwined Kelvin waves which twist and split into singly-charged vortices. Numerical experiments suggest that the onset of the decay is sensitive to small perturbations.
We have also shown that the twisted decay can be used
to generate isotropic turbulence, relatively free from large
scale fluctuations or fragmentation of the condensate, hence suitable for comparison with superfluid helium turbulence. Finally, an interesting follow-up study would be to assess how the proximity of the BEC edge and the presence of the thermal cloud would affect the Kelvin waves~\cite{Rooney2011}.

\begin{acknowledgments}
We acknowledge financial support from FAPESP (program CEPID), 
CNPq (program INCT) and EPSRC. 
G.D.T and P.E.S.T thank G.~Roati for technical support during the initial experiment runs. V.S.B thanks E.A.L.~Henn for the experimental support.
\end{acknowledgments}



\begin{thebibliography}{10}
\bibitem{Supp1}
The schematic imprinting process is shown in Fig.~1 
of the Supp. Mat. at [URL will be inserted by the publisher].
\bibitem{Supp2}
Examples of optical density profiles used for data analysis are shown
in Fig.~2 of the Supp. Mat. at  [URL will be inserted by the publisher].
\bibitem{Supp3}
The 1D optical density profile of the condensate taken before the decay 
of the multi-charged vortex are shown in Fig.~3 of the Supp. Mat. at
[URL will be inserted by the publisher].
\bibitem{Supp4}
See Fig.~4 of the Supp. Mat. at [URL will be inserted by the publisher].
\end{thebibliography}

\newpage
\centerline{\large \bf SUPPLEMENTARY MATERIAL}
\bigskip

\centerline{ {\bf 1. Topological phase imprinting} }
\bigskip

\begin{figure}[!ht]
\centering{
\includegraphics[scale = 0.5]{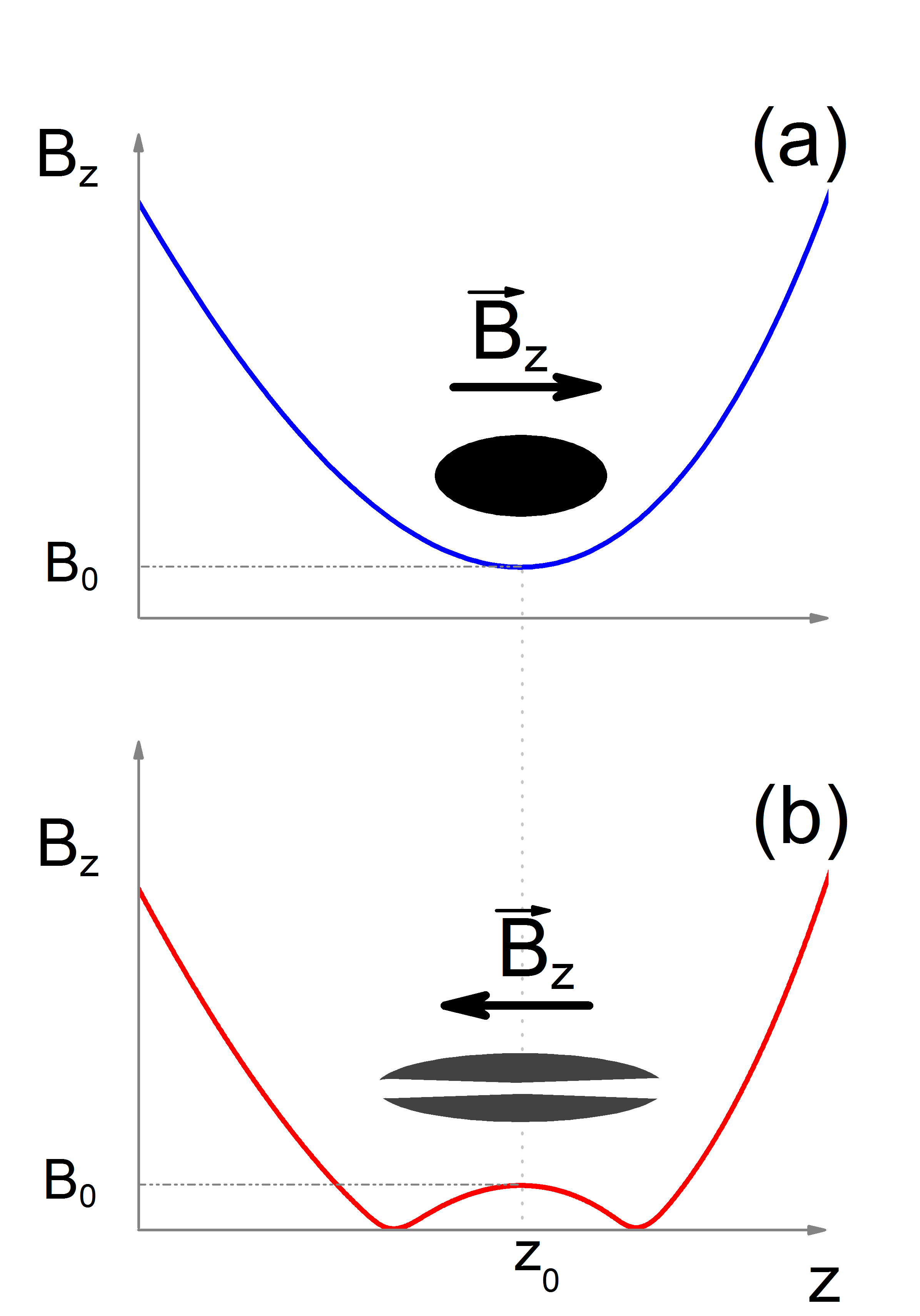}} 
\caption{{\bf Topological phase imprinting of a multi-charged vortex.} The
diagram summarizes the change in the axial magnetic field profile 
in the topological phase imprinting procedure used to 
nucleate multi-charged vortices in our BEC. 
(a) The initial (standard) magnetic field axial profile of our QUIC trap, 
held until the phase imprinting starts.
(b): The final magnetic field axial profile after the  
imprinting finishes, about $5~\unit{ms}$ later, 
just before it is switched off for the time-of-flight imaging.
}
\label{fig5}
\end{figure}

The multi-charged vortices are nucleated by adiabatically
reducing and inverting the initial trap bias field from 
$B_{0}\approx 0.5~\unit{G}$ 
down to $B_{0}\approx -0.5~\unit{G}$. The resulting magnetic 
field along the weak trap direction, $z$, axis is reversed during the process.  Fig.~\ref{fig5} presents the  general idea with the main details. During this adiabatic reversal process, the atomic spins are forced to follow the rotation taking place in the local field directions.  
The magnetic field reversal rate is slow enough ($0.1~\unit{G/ms}$) so that almost all the atomic spins adiabatically follow
the magnetic field directions, under our typical experimental conditions. Also, since the direction of the
rotation depends on the spatial position of each atom,
the atom acquires a different topological phase depending on its position. The spin reversal results in the phase winding in the condensate order parameter with $4\pi\hbar$ per atom, corresponding
to a $n=4$ charge vortex imprinted on the BEC. Finally, this process works well though it induces significant loss of atoms ($50\%$) during the zero crossing, due to the Majorana flips, which is consistent with previous studies..

\vskip 3cm
\newpage

\centerline{\bf 2. Profiles of multi- and singly-charged vortices}
\bigskip

\begin{figure}[!ht]
\centering{
\includegraphics[scale = 0.29]{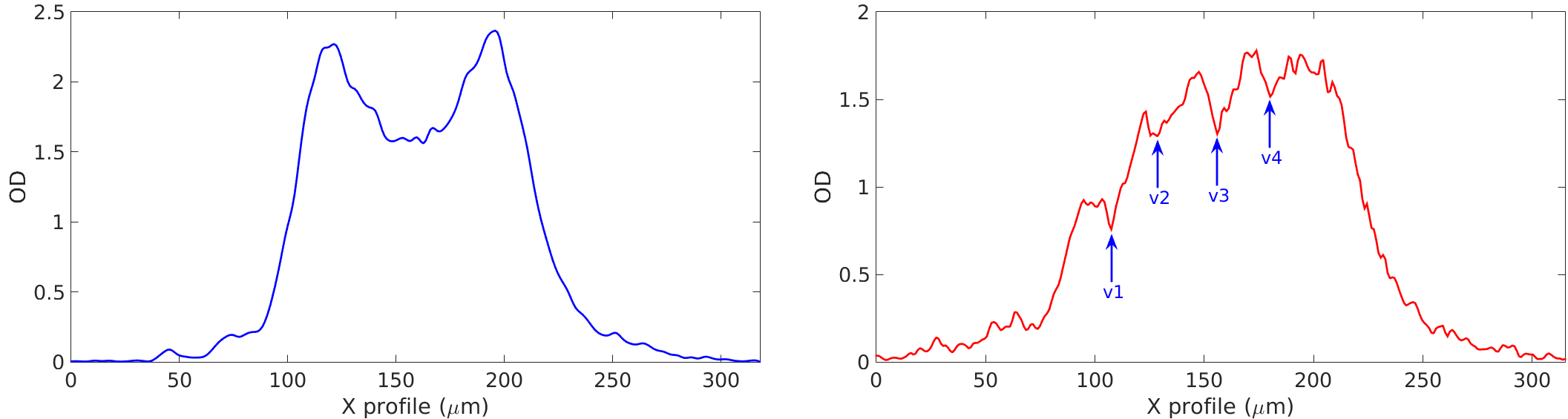}}
\caption{{\bf Profile of multi- and singly-charged vortices}
1D optical density profiles corresponding to the Fig.~1 of the article, for the multi-charged vortex (left) and the four singly-charged
vortices (right).
}
\label{fig6}
\end{figure}

In Fig.~\ref{fig6}, typical optical density 1D profiles, OD(z), 
are presented. On the left, in Fig.~\ref{fig6}(a), a $n=4$ multi-charged 
vortex with its large central depletion is shown; and, on the right, 
the four singly charged vortices resulting from its decay 
Fig.~\ref{fig6}(b) are presented.

\vskip 2cm

\newpage
\centerline{\bf 3. Profile of a vortex without Kelvin waves}
\bigskip

\begin{figure}[h!]
\centering{\includegraphics[scale = 0.65]{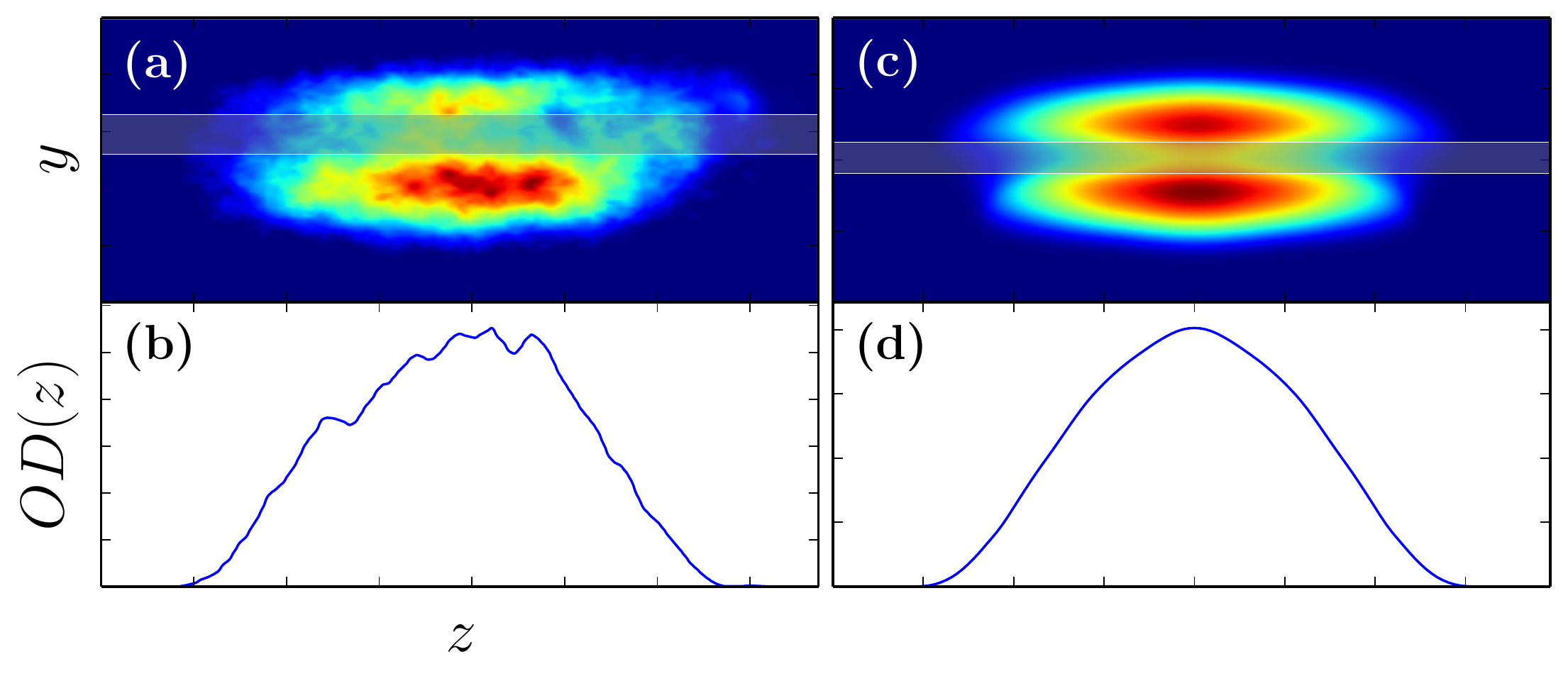}}\\
\caption{\textbf{Multicharged vortex as seen for short hold times ($2.2~\unit{ms}$):}
(a) experimental absorption images (false color); and (c) numerical 2D column density. (b) corresponding experimental OD $z-$profile; and (d) numerical central OD($z$) along the condensate $z-$axis (averaged along the narrow horizontal strips shown).
} 
\label{fig7}
\end{figure}

Fig.~\ref{fig7} shows the smooth
experimental (b) and numerical (d) 1D OZ(z) profiles of the central 
region of the condensate along the z-axis for a vortex without Kelvin
waves, corresponding to experimental (a) and numerical (b) false-color
absorption images. These smooth profiles must be compared to
the oscillating profiles for a vortex with Kelvin waves shown in
Fig.~\ref{fig2}(c,d,e,f) of the article.

\newpage
\centerline{ {\bf 4. Generation of quantum turbulence} }
\bigskip

\begin{figure}[h!]
\centering{\includegraphics[scale = 0.65]{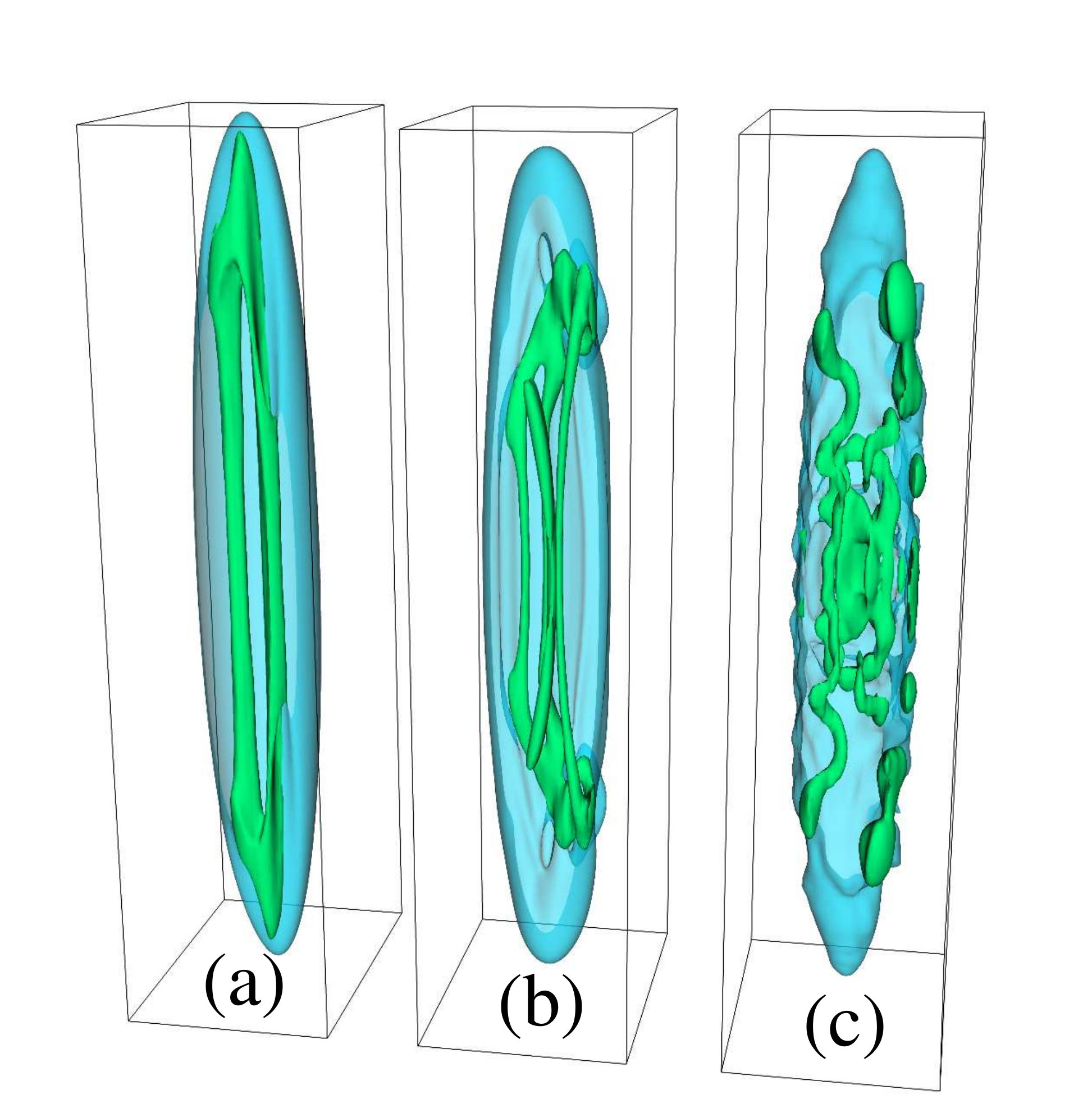}}\\
\caption{\textbf{Isodensity plots showing the evolution into turbulence} of two initial $n=2$ antiparallel vortices at $t=0$ ms (a), $t=21$ ms (b) and $t=53$ ms (c).
} 
\label{fig8}
\end{figure}

To exploit the twisted unwinding of the multi-charged vortices
as a technique to generate vortex tangles, relatively free of large density modulation, we numerically imprinted two anti-parallel, doubly quantized vortices, using the parameters listed below, at the initial condition, see Fig.\ref{fig8}(a). The vortices unwind, twist, move slightly forward due to the self-induced velocity field, and reconnect (Fig.\ref{fig8}(b)), 
generating a turbulent tangle, see Fig.\ref{fig8}(c).

{\bf Initial condition:} a doubly quantized ($n=2$)
vortex is imprinted in the xy plane at $(1.8\ell_r, 1.5 \ell_r)$,
and a double quantized antivortex is imprinted at
$(1.0 \ell_r, -1.3 \ell_r)$. All parameters as in the
numerical simulation described in the article.

{\bf Velocity statistics in the turbulent state:} the PDFs of the 
velocity components are
${\rm PDF}(v_i) \sim v_i^{\alpha_i}$ ($i=x,y,z)$ with 
$\alpha_x \approx -3.01$, $\alpha_y \approx -3.14$, 
$\alpha_z \approx -3.12$.
The normalized mean values of the Gaussian fits (i.e., mean
values
${\rm gPDF}(v_i)=(2 \pi \sigma^2_i)^{-1/2} 
\exp{[-(v_i-\mu_i)^2/(2 \sigma_i^2)]}$,
divided by their corresponding Gaussian widths) are marked as vertical dashed lines in Fig. 4. and
are respectively given by $v_i=-0.014,-0.210,-0.005$, with associated widths $\sigma_i=2.3,2.4,1.5$ and $\mu_i=-0.014,-0.210,-0.005$ (for $i=x,y,z$).

\end{document}